\documentclass[runningheads]{llncs}

\usepackage{eccv}

\usepackage{xcolor}
\usepackage{eccvabbrv}
\usepackage{multirow} 
\usepackage{graphicx}
\usepackage{booktabs}
\usepackage{lipsum}
\usepackage[accsupp]{axessibility}  

\usepackage{hyperref}

\usepackage{orcidlink}

\begin{document}

\title{OAPT: Offset-Aware Partition Transformer for Double JPEG Artifacts Removal} 

\titlerunning{OAPT}

\author{Qiao Mo$^{1,2,\star}$\orcidlink{0009-0002-7250-3056} \and
Yukang Ding$^{2,\star}$\orcidlink{0009-0003-9716-0798} \and
Jinhua Hao$^{2,\dagger}$\orcidlink{0000-0003-4571-2063} \and 
Qiang Zhu$^{1}$\orcidlink{0009-0002-8034-3275} \and
Ming Sun$^{2}$\orcidlink{0000-0002-5948-2708} \and
Chao Zhou$^{2}$\orcidlink{0000-0003-2969-3042} \and
Feiyu Chen$^{1}$\orcidlink{0000-0002-0928-6899} \and
Shuyuan Zhu$^{1,\dagger}$
}

\authorrunning{Q.~Mo, Y.~Ding et al.}

\institute{University of Electronic Science and Technology of China \\
\email{eezsy@uestc.edu.cn} \\
\and 
Kuaishou Technology
 \\
\email{\{moqiao, dingyukang, haojinhua\}@kuaishou.com}
} 
\maketitle

\renewcommand{\thefootnote}{}
\footnotetext{$^\star$ Equal contribution. ~$^\dagger$ Corresponding author.}

\begin{abstract}
  Deep learning-based methods have shown remarkable performance in single JPEG artifacts removal task. However, existing methods tend to degrade on double JPEG images, which are prevalent in real-world scenarios. 
  To address this issue, we propose \textbf{O}ffset-\textbf{A}ware \textbf{P}artition \textbf{T}ransformer for double JPEG artifacts removal, termed as \textbf{OAPT}. We conduct an analysis of double JPEG compression that results in up to four patterns within each $8\times8$ block and design our model to cluster the similar patterns to remedy the difficulty of restoration.
  Our OAPT consists of two components: compression offset predictor and image reconstructor. Specifically, the predictor estimates pixel offsets between the first and second compression, which are then utilized to divide different patterns.
  The reconstructor is mainly based on several Hybrid Partition Attention Blocks (HPAB), combining vanilla window-based self-attention and sparse attention for clustered pattern features. 
  Extensive experiments demonstrate that OAPT outperforms the state-of-the-art method by more than 0.16dB in double JPEG image restoration task. Moreover, without increasing any computation cost, the pattern clustering module in HPAB can serve as a plugin to enhance other transformer-based image restoration methods. The code will be available at \url{https://github.com/QMoQ/OAPT.git}.
\end{abstract}

\section{Introduction}
\label{sec:intro}
JPEG~\cite{jpeg} is the most widely-used algorithm for image compression~\cite{jpegsurvey1,jpegsurvey2,jpegsurvey4,jpegsurvey5}. Renowned for its simplicity and efficiency, JPEG is extensively applied in bandwidth-constrained scenarios. 
It splits images into $8\times8$ blocks, then applies discrete cosine transform (DCT~\cite{dct}) and quantization operations to each block, resulting in lossy compression. 
The degree of compression can be adjusted by the JPEG Quality Factor (QF), ranging from 0 to 100, where smaller QFs indicate more aggressive compression~\cite{jpeg, jpegsurvey2,jpegsurvey4}. While heavy compression can reduce storage space, it comes at the cost of compromised image quality, leading to noticeable annoying blocky artifacts~\cite{jpegsurvey3,jpegsurvey5}. 
The removal of single JPEG artifacts has attracted attention~\cite{idcn, swinir, rnan, dmcnn, cascnn} due to its practical usage.
Over the last decade, numerous convolutional neural network (CNN) based methods~\cite{arcnn, dcnn, drunet, rdn, qgac, fbcnn, rnan,cpga,realblur} have been proposed to address the ill-posed image restoration tasks, achieving notable success in academia. In recent years, transformer-based methods~\cite{swinir, grl, art,cassr,xpsr} have surpassed CNN-based approaches by leveraging the promising self-attention capability~\cite{vit}.

\begin{figure*}[!tb]
\centering
\includegraphics[width=\textwidth]{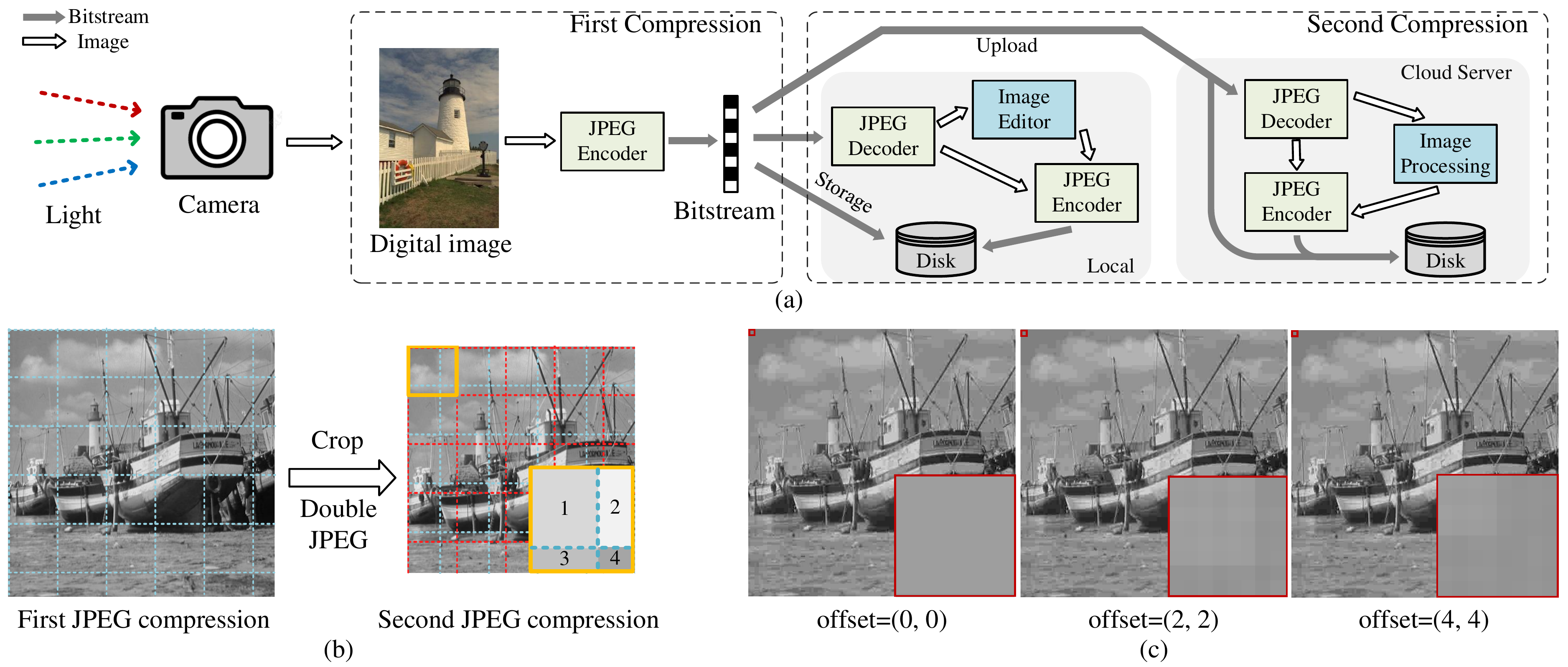}
\caption{The demonstration of double compression.}
\label{fig.double_JPEG}
\end{figure*}

Compared with single compression, double JPEG compression is more common in real-world scenarios, as images on the internet always undergo multiple compression cycles~\cite{fbcnn,f10,f29,f40}. 
For example, as shown in \cref{fig.double_JPEG}(a), a newly-taken photo, which has already gone through the first-time compression for being stored locally in the disk of cameras, will be conducted the second-time compression after image editing, like cropping, or uploading to the cloud server of social media, making it a double-compressed image. With the recurrence of the process mentioned above, multi-compressed images are generated naturally.
However, most previous methods are trained on single compression data and only account for the range of QFs, encountering a substantial performance decline in double JPEG image restoration, as emphasized in~\cite{fbcnn}. 
FBCNN~\cite{fbcnn}, as a pioneering effort for double JPEG artifacts removal, estimates dominant QF in double compression or gets fine-tuned on synthesized double JPEG images. Though being effective in both single and double compression artifacts reduction, it is not explicitly designed for the characteristics of double JPEG compression.
By analyzing the occurrence of double JPEG images, we found that besides various combination of QFs, there is obvious compression shift in double JPEG images.
With the pixel-level offsets, non-aligned compression results in no more than four kinds of patterns within each $8\times8$ block from the second compression, as illustrated in \cref{fig.double_JPEG}(b).
It brings distinct borders and varying pattern characteristics in $8\times8$ blocks, as shown in \cref{fig.double_JPEG}(c). We observed that the non-aligned compression image restoration is a much tougher task by testing DnCNN~\cite{dcnn} in various double compression scenarios. A noticeable decrease of PSNR appeared in non-aligned compression when compared to aligned compression.

To address the non-aligned compressed image restoration, we attempt to obtain the compression offsets and aggregate the same patterns together for better restoration. Drawing inspiration from the partition operation in Swin Transformer~\cite{swin, swinir}, a novel partition strategy is proposed for the same positional patterns in $8\times8$ blocks.
Based on it, we design an \textbf{O}ffset-\textbf{A}ware \textbf{P}artition \textbf{T}ransformer, namely OAPT. 
OAPT consists of two main components: a CNN-based compression offset predictor and a transformer-based image reconstructor.
In detail, the compression offset predictor estimates pixel offsets between two JPEG compressions, allowing for effective clustering of patterns that exhibit similar compression effects at the same position in each $8\times8$ block across the entire image.
The image reconstructor, composed of Hybrid Partition Attention Blocks (HPAB), incorporates window-based self-attention and sparse attention for clustered pattern features. The alternating attention and pattern clustering mechanism can enhance the robustness of double compression artifacts removal. 
Experiments show OAPT achieves the best performance and relatively has less parameters than many methods. 
Moreover, our pattern clustering module in HPAB can be designed as a plugin, improving the performance of other transformer-based methods without introducing additional parameters.

To conclude, the contribution of our work includes:
\begin{itemize}
    \item We design a novel offset-aware partition transformer (OAPT). OAPT estimates the pixel-level offsets in double compression and uses them to adaptively provide hybrid attention for both dense features and clustered features by HPABs. It reduces difficulty in non-aligned compression artifacts removal.
    \item The pattern clustering module in HPAB can be used as a plugin on the transformer-based methods, which brings improvement to the double compression image restoration without extra computation cost and parameters.
    \item Experiments show that our OAPT obtains great performance compared with previous CNN-based and transformer-based methods. It outperforms the state-of-the-art method~\cite{fbcnn} by 0.16dB in double JPEG image restoration.
\end{itemize}

\section{Related Work}
\label{sec:relatedwork}
\subsubsection{Single JPEG Image Restoration}
Deep learning-based methods for JPEG image restoration have witnessed significant advancements over the last decade. ARCNN~\cite{arcnn} is the first to introduce a deep learning network for restoring JPEG images. 
~\cite{dcnn,rdn,mdsr} achieve improvement by incorporating residual learning~\cite{resnet} and batch normalization~\cite{bn}.
Local self-similarity and non-local similarity are proved to be significant for image restoration and used for enhancement in~\cite{rnan,panet,dagl}. 
Meanwhile,~\cite{drunet,fu,qgac,ircnn,mwcnn,tnrd,rich-bsr} explore various other priors to improve performance of image restoration. 
Moreover, dual-domain processing~\cite{strrn,idcn,dmcnn,jpegdeartdct} is adopted to extract diverse types of information.

The efficacy of the self-attention mechanism~\cite{vit} and its ability to handle long-range dependencies have propelled transformer-based image restoration methods~\cite{uformer, ipt, swinir, hat, grl, restormer, cat, art} surpassing previous CNN-based approaches. IPT~\cite{ipt} develops a pretrained model utilizing the powerful representation ability of transformer for low-level computer vision tasks. Drawing on insights from~\cite{swin}, SwinIR~\cite{swinir} demonstrates promising performance across various image restoration tasks, including JPEG artifacts removal. GRL~\cite{grl} presents an effective architecture by modeling image hierarchies at global, regional, and local scales. Additionally, ART~\cite{art}, coupled with a sparse attention module, can capture a broader receptive field, thereby outperforming existing methods across multiple image restoration tasks.

\subsubsection{Double JPEG Artifacts Removal}
The realm of double JPEG compression has long captivated researchers, particularly in the field of detecting and localizing image forgery. Research studies~\cite{f14, f8} have illuminated substantial distinctions between single and double JPEG compression. In~\cite{f8, f3, f29, f4, djd6}, double JPEG compression was categorized into two classes, aligned and non-aligned double compression, for in-depth researching. Learning-based methods~\cite{djd1, djd2, djd3, djd4, djd5} leverage information from both spatial and DCT domains to discern forged regions. In~\cite{f16, f10, f44, f40, f31, f2}, due to the difference in quantization and compression, the initial compression's quantization information is exploited to distinguish areas subjected to single and double JPEG compression.

Despite the apparent progress in double compression detection, methods for restoring double-compressed images remain scarce. FBCNN~\cite{fbcnn} stands out as the algorithm dedicated to image restoration that encompasses non-aligned double-compressed images. FBCNN offers two strategies to address the problem of double JPEG image restoration. The first involves rectifying the estimated Quality Factor (QF) to align with the dominant one. The second strategy entails training the network with double JPEG compression images, utilizing the QF predictor in an unsupervised manner. However, both approaches do not fully exploit the characteristics of double compression, especially the different patterns caused by pixel offsets, shown in \cref{fig.double_JPEG}(b). 
It has motivated us to explore a more practical method for tackling this challenging image restoration task.

\section{Method}
\label{sec:method}
\subsection{Motivation}
\label{sec:method_overview}
As highlighted in~\cite{fbcnn} that existing methods trained solely with single compression data struggle when applied to double compression image restoration, we performed the experiment using DnCNN~\cite{dcnn} which was trained with grayscale double JPEG images with various QFs and offsets. 
We generated two sets of grayscale images from LIVE1~\cite{live1} by the data degradation mentioned in Sec.\ref{sec/exp_setup}, which are \textbf{aligned} double JPEG images with $\text{offset=(0, 0)}$ and \textbf{non-aligned} double JPEG images with $\text{offset=(4, 4)}$.
We calculate the $\Delta$PSNR between the low-quality images and their corresponding enhanced images processed by DnCNN under $(QF_1, QF_2)=(30, 50)$ and $(QF_1, QF_2)=(50, 30)$. The results of $\Delta$PSNR with aligned double compressed images are respectively 2.06 dB and 3.24dB, while the results of $\Delta$PSNR with non-aligned compressed images are 1.66 dB and 1.64 dB.
Notably, the performance on non-aligned double compressed images is inferior to that on aligned double compressed images. Learning from that, we can draw a similar conclusion with ~\cite{fbcnn} that restoration of non-aligned double JPEG images is more complex and difficult.

In \cref{fig.double_JPEG}(b), it delineates the shifting offsets in double compression cause four patterns in each $8\times 8$ coding unit. The four parts locate in the same $8 \times 8$ block in the second compression, while belonged to 4 different $8 \times 8$ blocks in the first compression. 
So after the second compression, four parts in the same $8\times 8$ block are degraded by different kinds of block effects and compression patterns.
It inspired us that the potential simplification of double JPEG image restoration might be individually processing the clustered patches of the same patterns. Considering the sparse clustered patches will harm the locality of images, unlike transformer-based models, CNN-based methods which depend its modeling solely on the relative position are not likely to benefit from it~\cite{shuffleformer}. Meanwhile, transformer-based methods usually perform window partition before self-attention as a pre-processing to reduce the computation cost~\cite{swinir,hat,grl}, which is similar to the pattern clustering. So we introduce the \textbf{O}ffset-\textbf{A}ware \textbf{P}artition \textbf{T}ransformer, denoted as OAPT. It is designed to estimate compression offsets and dynamically provide hybrid attention tailored to different patterns.

\subsection{Network Architecture}

\begin{figure}[t]
	\centering
	\includegraphics[width=\textwidth]{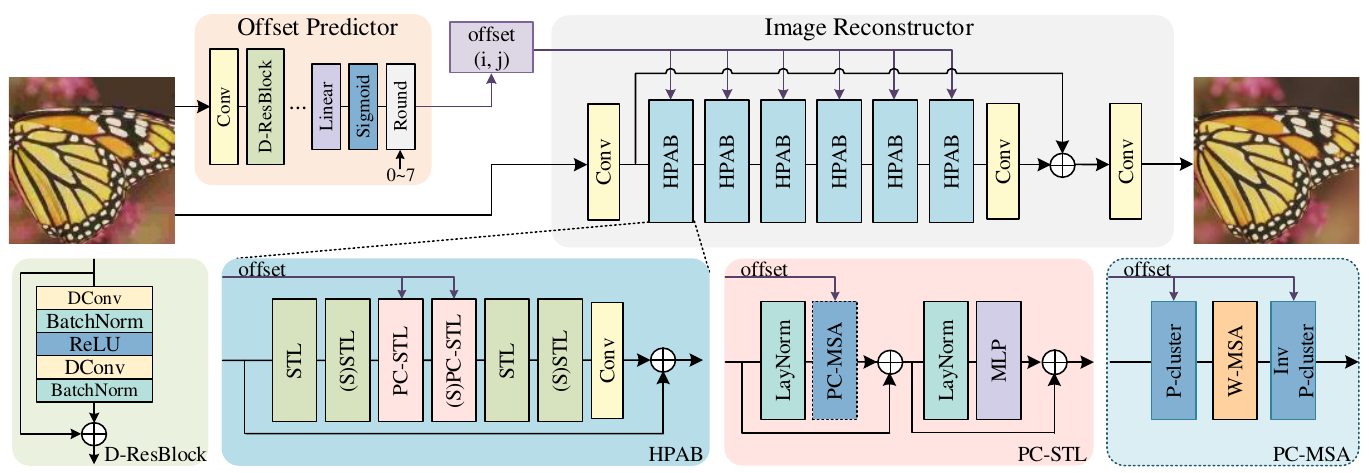}
	\caption{The architecture of Offset-Aware Partition Transformer(OAPT). In PC-MSA, \textbf{P-cluster} represents Pattern clustering and \textbf{Inv P-cluster} stands for the inverse operation of pattern clustering. \textbf{(S)} means equipping with window shifting operation.}
	\label{fig.overview}
\end{figure}

The architecture of OAPT is depicted in \cref{fig.overview}. It comprises two main components: a ResNet-based compression offset predictor and a transformer-based image reconstructor. The compression offset predictor is tasked with estimating the compression shift in both rows and columns, while the reconstructor aims to restore images using the predicted compression offsets.

\subsubsection{Compression Offset Predictor}
As analyzed in Sec.\ref{sec:method_overview}, the shifting manifests in $8\times 8$ blocks which are uniform and periodic, indicating that the offsets of rows and columns range from 0 to 7. During image degradation, we randomly remove the first $i$ rows and $j$ columns in the second compression to simulate non-aligned compression, with the resulting offsets labeled as $(i, j)$. Detailed information on synthesizing double compressed images is provided in Sec.\ref{sec/exp_setup}.

As depicted in \cref{fig.overview}, the compression offset predictor relies on ResNet-18~\cite{resnet}. It takes low-quality images as inputs and estimates the offsets of rows and columns. Differing from the resconstructor, the predictor's input is only a $44\times44$ patch from the top-left corner of the input image where JPEG starts to split the image into blocks and conducts compression. To minimize the number of parameters, we replace all convolution layers in residual blocks with depthwise separable convolution~\cite{dconv} layers, denoting these modified blocks as D-Resblocks. Additionally, we introduce an extra linear layer, sigmoid activation function, and round operation to generate an output consisting of two integers between 0 and 7, shown in \cref{offset_pred}, 
\begin{equation} \label{offset_pred}
\begin{aligned}\relax
[\hat{r}, \hat{c}] = \text{Round}(\text{Sigmoid}([r', c']) \times 7),
\end{aligned}
\end{equation}
where $r'$ and $c'$ are the output of the last linear layer, while $\hat{r}$ and $\hat{c}$ are the final predicted offsets of rows and columns. Leveraging a compact CNN network and the reduced input size, the computation cost of the predictor is substantially lower than that of the reconstructor.
We optimize the offset predictor by minimizing
\begin{equation} \label{offset_loss}
\begin{aligned}
\mathcal{L}_{\textit{offset}} = \| \hat{r} - r \|_{1} + \|  \hat{c} - c \|_{1}, 
\end{aligned}
\end{equation}
where $r$ and $c$ are the ground-truth offsets of rows and columns.

\begin{figure}[t]
	\centering
	\includegraphics[width=\textwidth]{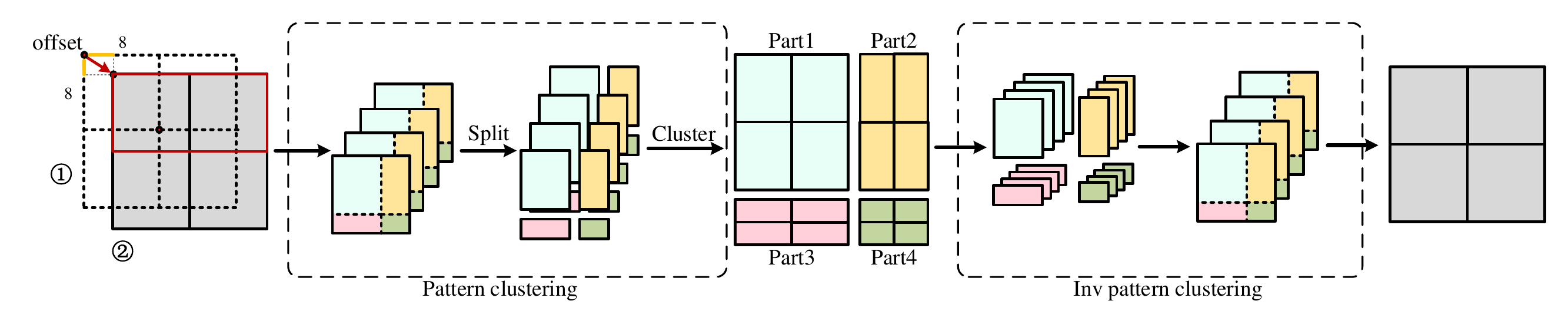}
	\caption{The demonstration of pattern clustering module.}
	\label{fig.pattern}
\end{figure}

\subsubsection{Hybrid Partition Attention Blocks based Image Reconstructor}
Inspired by the window partition operation in~\cite{swin}, we design a new partition operation for our hybird attention to match double JPEG image restoration. With the estimated offsets from the predictor, we can split every $8\times 8$ block to four parts and cluster the same positional patterns for the following self-attention. Our image reconstructor is mainly composed of this new hybrid attention.

In detail, the image reconstructor are based on three parts: one convolution layer for shallow feature extraction, serials of transformer-based modules for deep feature extraction and one convolution layer for high-quality image reconstruction. The part for deep feature extraction is the core of OAPT, which consists of \textbf{H}ybrid \textbf{P}artition \textbf{A}ttention \textbf{B}locks (HPAB) and one convolution layer.

The HPAB is a residual block with one convolution layer, four Swin Transformer Layers (STL) and two Pattern Clustering-based Swin Transformer Layers (PC-STL), as illustrated in \cref{fig.overview}. The HPAB can provide vanilla window-based self-attention~\cite{swin} and sparse attention. The STL proposed in~\cite{swinir} is focused on the dense attention for local continuous features within every window. The PC-STL is equipped with a pattern clustering-based multi-head self-attention (PC-MSA). The pattern clustering module takes the estimated offsets to divide every $8\times8$ block into four patterns and clusters patterns in the same position for self-attention respectively, which is shown in \cref{fig.pattern}. After clustering, the PC-STL achieves sparse pattern attention and then reverses the clustered patterns to their original positions. Seemingly being similar to the sparse attention in ART~\cite{art}, ours decomposes the input feature into 4 relatively sparse patches according to the offsets for better extracting information in the same patterns, while the purpose of sparse attention with uniform sampling in ART is to enlarge the receptive field size. Meanwhile, our module advantages the attention module of ART in less parameters and computation cost.

The PC-STL is expressed as \cref{pc-stl}, 
\begin{equation}\label{pc-stl} %
\centering
\begin{aligned}
X'=\text{PC-MSA}(&\text{LayerNorm}(X), \textit{offset}) + X, \\
Y = \text{MLP}(&\text{LayerNorm}(X')) + X',
\end{aligned}
\end{equation}

where $X$ stands for the input feature of PC-STL, $\textit{offset}$ is the estimated offsets from the predictor and $Y$ is the output of PC-STL. In detail, PC-MSA can be expressed as below:
\begin{equation} \label{pc-msa} %
\centering
\begin{aligned}
\left[x_1, x_2, x_3, x_4\right] =& \text{PC}(X_{\text{LN}}, \textit{offset}), \\
\hat{X} = \text{invPC}(\text{W-MSA}&([x_1, x_2, x_3, x_4]), \textit{offset}),
\end{aligned}
\end{equation}

where \text{PC} denotes pattern clustering module, $x_i$ represents the clustered parts of the similar pattern after pattern clustering, \text{invPC} denotes the inverse operation of pattern clustering and $X_{LN}$ denotes the input feature of PC-MSA. \text{W-MSA} represents the typical window-based multi-head self-attention~\cite{vit} for each clustered part separately, which can be also formulated as: 
\begin{equation} \label{sa} 
\centering
\begin{aligned}
\text{Attention}(Q,K,V)=\text{Softmax}(QK^T/\sqrt{d}+B)V,
\end{aligned}
\end{equation}
where $Q$, $K$, $V$ are respectively the query, key, value from the linear projecting of input $x_i$, $B$ is the learnable relative positional encoding and $d$ denotes the dimension size of each token.

As the Charbonnier loss~\cite{cbloss} is effective in image restoration, we optimize the reconstructor by minimizing the pixel loss:
\begin{equation} \label{rec_loss}
\mathcal{L}_{\textit{rec}}=\sqrt{\|\hat{I}-I\|^2+\epsilon^2},
\end{equation}
where $\epsilon$ is a constant value of $10^{-3}$, $\hat{I}$ is the reconstructed image by the reconstructor, and $I$ is the corresponding ground-truth high-quality image. 

\subsection{Pattern Clustering Plugin Module for Transformer based Methods}
The proposed pattern clustering in HPAB is a robust mechanism, capable of functioning as a plugin module for other transformer-based methods without introducing additional parameters or computational overhead. Transformer-based networks can easily incorporate the pattern clustering as a pre-processing before self-attention to enhance non-aligned double compressed image restoration. And the inverse pattern clustering can also be incorporated as a post-processing for rearranging the clustered patches to their original position after going through self-attention. So our pattern clustering plugin module is composed of pattern clustering operation and inverse pattern clustering operation. We conduct the experiment of implementing pattern clustering mechanism on HAT-S~\cite{hat} and the details are demonstrated in \cref{sec:exp_of_plugin}. The results show that this cheap plugin module not only contributes improvement to HAT-S, but also expand the receptive field without extra parameters or computation complexity.

\section{Experiments} 
\subsection{Experimental Setup}
\label{sec/exp_setup}
\subsubsection{Datasets and Degradation Methods.} For datasets, we use the Y channel of YCbCr space for grayscale image datasets, and the RGB channels for color image datasets. Same as~\cite{qgac,fbcnn}, we train our OAPT network on DIV2K~\cite{div2k} and Flickr2K~\cite{flick}. 
We follow the degradation setting in~\cite{fbcnn}, the degradation model can synthesize both aligned and non-aligned double JPEG images via
\begin{equation} \label{double_JPEG}
\begin{aligned}
y = \text{JPEG}(\text{Shift}_{i,j}(\text{JPEG}(x, \text{QF}_1)), \text{QF}_2).
\end{aligned}
\end{equation}
\noindent
Among them, $x$ presents the high-quality image, $y$ denotes the double JPEG compressed image. $\text{QF}_1$ and $\text{QF}_2$ are quality factors of the first and second compression, which are both randomly sampled from 5 to 95. The $\text{Shift}$ is the random removal of the first $i$ rows and $j$ columns in the second compression, where $0 <= i,j <= 7$. Especially, we only train a \textbf{single} model to cover all the range of QFs and the offsets of pixels in double compression. 

\subsubsection{Training Details.} For the offset predictor, the number of D-Resblocks is 8, and the max channel is 512. For the image reconstructor, we set the HPAB number, channel number and window size to 6, 180 and 7, respectively. The size of randomly extracted input image patches is $224 \times 224$ and the batchsize is 4. 
To optimize the parameters, we adopt Adam solver~\cite{adam} and initialize the learning rate as $2e^{-4}$. 
We pretrain the offset predictor first and train our OAPT model with freezing the parameters of the offset predictor on 4 Nvidia V100 GPUs.
As the main backbone is similar with SwinIR, we initialized the image reconstructor of ours with the pretrained model of SwinIR, and fine-tune it on double compression datasets.

\begin{table*}[tb]\scriptsize 
\center
\begin{center}
\caption{Quantitative comparison (average PSNR/SSIM/PSNR-B) with state-of-the-art methods on \textbf{grayscale} double JPEG images and computation cost. \textcolor{red}{Best} performance is noted in \textcolor{red}{Red}. ($\text{QF}_{1}$, $\text{QF}_{2}$, $i$, $j$) in Type means the combination types of double compression. $\text{QF}_{1}$ and $\text{QF}_{2}$ are quality factors of the first and the second compression, $i$ and $j$ denote the pixel-level shifting offset $(i, j)$ between two compressions.}
\label{tab:sota_table}
\resizebox{\columnwidth}{!}{
\begin{tabular}{|c|c|c|c|c|c|c|c|c|}
\hline
Dataset & Type & DnCNN~\cite{dcnn} & RNAN~\cite{rnan} & FBCNN~\cite{fbcnn} & HAT-S~\cite{hat} & ART~\cite{art} & SwinIR~\cite{swinir} & OAPT(Ours) \\
\hline
\hline
\multirow{8}{*}{Classic5\cite{class5}}
& (30, 30, 4, 4) & 31.68/0.8603/31.41 & 31.89/0.8652/31.59 & 32.12/0.8686/31.89 & 32.28/0.8707/31.99 & 32.29/0.8716/31.96 & 32.26/0.8703/31.95 & \textcolor{red}{32.32}/\textcolor{red}{0.8718}/\textcolor{red}{32.07}  \\
& (50, 50, 4, 4) & 33.22/0.8903/32.90 & 33.46/0.8943/33.09 & 33.70/0.8974/33.33 & 33.85/0.8985/33.44 & 33.86/0.8990/33.37 & 33.80/0.8980/33.40 & \textcolor{red}{33.87}/\textcolor{red}{0.8992}/\textcolor{red}{33.51}  \\
& (30, 50, 4, 4) & 32.30/0.8737/32.08 & 32.51/0.8777/32.25 & 32.74/0.8810/32.56 & 32.91/0.8827/32.67 & 32.93/0.8834/32.69 & 32.90/0.8830/32.68 & \textcolor{red}{33.02}/\textcolor{red}{0.8851}/\textcolor{red}{32.77}  \\
& (50, 30, 4, 4) & 32.31/0.8731/31.99 & 32.57/0.8784/32.21 & 32.81/0.8822/32.38 & 32.96/0.8835/32.52 & \textcolor{red}{33.02}/\textcolor{red}{0.8850}/\textcolor{red}{32.55} & 32.95/0.8833/32.46 & 32.97/0.8840/32.54 \\
& (30, 30, 0, 4) & 31.83/0.8652/31.55 & 32.05/0.8703/31.74 & 32.30/0.8736/31.99 & 32.46/0.8754/32.13 & 32.47/0.8764/32.08 & 32.43/0.8753/32.05 & \textcolor{red}{32.50}/\textcolor{red}{0.8768}/\textcolor{red}{32.18}  \\
& (50, 50, 0, 4) & 33.34/0.8937/33.03 & 33.57/0.8975/33.22 & 33.85/0.9007/33.38 & 33.99/0.9017/33.52 & 33.98/0.9020/33.43 & 33.94/0.9014/33.43 & \textcolor{red}{34.02}/\textcolor{red}{0.9026}/\textcolor{red}{33.57}  \\
& (30, 50, 0, 4) & 32.44/0.8776/32.18 & 32.65/0.8819/32.37 & 32.93/0.8855/32.64 & 33.07/0.8870/32.80 & 33.10/0.8878/32.82 & 33.06/0.8872/32.81 & \textcolor{red}{33.16}/\textcolor{red}{0.8889}/\textcolor{red}{32.89}  \\
& (50, 30, 0, 4) & 32.44/0.8769/32.10 & 32.69/0.8821/32.32 & 32.94/0.8860/32.38 & 33.11/0.8872/32.61 & \textcolor{red}{33.14}/\textcolor{red}{0.8883}/\textcolor{red}{32.62} & 33.09/0.8869/32.53 & 33.11/0.8874/32.61 \\

\hline
\hline
\multirow{8}{*}{LIVE1\cite{live1}} 
& (30, 30, 4, 4) & 31.58/0.8789/31.30 & 31.78/0.8836/31.47 & 31.94/0.8859/31.73 & 32.05//0.8875/31.79 & 32.07/0.8885/31.76 & 32.04/0.8871/31.76 & \textcolor{red}{32.10}/\textcolor{red}{0.8889}/\textcolor{red}{31.88} \\
& (50, 50, 4, 4) & 33.28/0.9120/32.91 & 33.51/0.9161/33.06 & 33.70/0.9188/33.34 & 33.81/0.9199/33.40 & 33.83/0.9204/33.33 & 33.80/0.9195/33.38 & \textcolor{red}{33.88}/\textcolor{red}{0.9210}/\textcolor{red}{33.52}  \\
& (30, 50, 4, 4) & 32.26/0.8939/32.04 & 32.42/0.8976/32.18 & 32.64/0.9004/32.49 & 32.74/0.9018/32.56 & 32.76/0.9023/32.51 & 32.77/0.9023/32.59 & \textcolor{red}{32.86}/\textcolor{red}{0.9043}/\textcolor{red}{32.70}  \\
& (50, 30, 4, 4) & 32.26/0.8927/31.88 & 32.50/0.8980/32.05 & 32.69/0.9012/32.24 & 32.77/0.9019/32.27 & \textcolor{red}{32.83}/\textcolor{red}{0.9033}/32.30 & 32.78/0.9016/32.25 & 32.80/0.9025/\textcolor{red}{32.34}  \\
& (30, 30, 0, 4) & 31.80/0.8849/31.51 & 31.97/0.8892/31.62 & 32.16/0.8914/31.76 & 32.26/0.8930/31.86 & 32.28/0.8939/31.84 & 32.25/0.8930/31.84 & \textcolor{red}{32.35}/\textcolor{red}{0.8946}/\textcolor{red}{31.97}  \\
& (50, 50, 0, 4) & 33.51/0.9168/33.17 & 33.72/0.9205/33.29 & 33.94/0.9229/33.37 & 34.05/0.9239/33.51 & 34.06/0.9245/33.43 & 34.04/0.9238/33.48 & \textcolor{red}{34.13}/\textcolor{red}{0.9252}/\textcolor{red}{33.60}  \\
& (30, 50, 0, 4) & 32.46/0.8994/32.23 & 32.64/0.9032/32.37 & 32.88/0.9061/32.51 & 32.97/0.9073/32.65 & 33.02/0.9082/32.69 & 32.99/0.9077/32.70 & \textcolor{red}{33.08}/\textcolor{red}{0.9095}/\textcolor{red}{32.77}  \\
& (50, 30, 0, 4) & 32.45/0.8976/32.09 & 32.69/0.9028/32.23 & 32.88/0.9061/32.29 & 32.98/0.9065/32.38 & \textcolor{red}{33.03}/\textcolor{red}{0.9078}/32.40 & 32.98/0.9062/32.37 & 33.00/0.9071/\textcolor{red}{32.45}  \\
\hline
\hline

\multirow{8}{*}{BSDS500\cite{bsds500}}
& (30, 30, 4, 4) & 31.48/0.8757/31.16 & 31.64/0.8800/31.26 & 31.81/0.8822/31.56 & 31.87/0.8833/31.57 & 31.89/0.8843/31.51 & 31.87/0.8828/31.54 & \textcolor{red}{31.93}/\textcolor{red}{0.8847}/\textcolor{red}{31.66}  \\
& (50, 50, 4, 4) & 33.18/0.9102/32.76 & 33.36/0.9139/32.84 & 33.54/0.9163/33.15 & 33.63/0.9171/33.17 & 33.64/0.9176/32.05 & 33.62/0.9166/33.14 & \textcolor{red}{33.68}/\textcolor{red}{0.9182}/\textcolor{red}{33.27} \\
& (30, 50, 4, 4) & 32.14/0.8910/31.87 & 32.27/0.8943/31.96 & 32.49/0.8970/32.31 & 32.55/0.8980/32.34 & 32.57/0.8984/32.29 & 32.59/0.8985/32.38 & \textcolor{red}{32.67}/\textcolor{red}{0.9006}/\textcolor{red}{32.47}  \\
& (50, 30, 4, 4) & 32.16/0.8902/31.72 & 32.37/0.8950/31.82 & 32.53/0.8977/32.03 & 32.59/0.8983/32.00 & \textcolor{red}{32.64}/\textcolor{red}{0.8996}/32.00 & 32.59/0.8977/31.97 & 32.61/0.8988/\textcolor{red}{32.08}  \\
& (30, 30, 0, 4) & 31.69/0.8817/31.31 & 31.84/0.8857/31.38 & 32.03/0.8876/31.55 & 32.09/0.8888/31.60 & 32.11/0.8898/31.52 & 32.08/0.8887/31.54 & \textcolor{red}{32.17}/\textcolor{red}{0.8907}/\textcolor{red}{31.71} \\
& (50, 50, 0, 4) & 33.41/0.9151/32.92 & 33.57/0.9183/32.99 & 33.78/0.9204/33.06 & 33.85/0.9212/33.15 & 33.86/0.9218/33.01 & 33.85/0.9211/33.08 & \textcolor{red}{33.93}/\textcolor{red}{0.9227}/\textcolor{red}{33.24}  \\
& (30, 50, 0, 4) & 32.36/0.8966/32.02 & 32.52/0.9000/32.11 & 32.74/0.9027/32.27 & 32.79/0.9036/32.37 & 32.84/0.9045/32.37 & 32.82/0.9040/32.41 & \textcolor{red}{32.91}/\textcolor{red}{0.9060}/\textcolor{red}{32.48}  \\
& (50, 30, 0, 4) & 32.36/0.8952/31.85 & 32.56/0.8999/31.94 & 32.74/0.9026/31.99 & 32.80/0.9031/32.03 & \textcolor{red}{32.85}/\textcolor{red}{0.9043}/32.01 & 32.81/0.9027/32.00 & 32.82/0.9035/\textcolor{red}{32.10}  \\
\hline
\hline
Params. & $\times$ & 0.67M & 8.96M & 71.92M & 9.24M & 16.14M & 11.49M & 12.96M \\ 
\hline
\hline
MACs & $\times$ & 17.08G & 193.98G & 71.21G & 227.14G & 415.51G & 293.42G & 293.60G \\ 
\hline
\end{tabular}}
\end{center}
\end{table*}

\begin{table}[tb] 
  \centering
  \caption{Quantitative comparison (average PSNR/SSIM/PSNR-B) on \textbf{color} double JPEG images.}
  \resizebox*{1.01\columnwidth}{!}{
    \begin{tabular}{|c|c|c|c|c|c|c|c|}
    \hline
    Dataset   & Method & (30, 30, 4, 4) & (50, 50, 4, 4)  & (30, 50, 4, 4) & (30, 30, 0, 4) & (50, 50, 0, 4)  & (30, 50, 0, 4) \\
    \hline
    \hline
    \multirow{4}{*}{ICB\cite{icb}}
                        & JPEG &32.18/0.8077/31.51 &33.47/0.8339/32.87 &32.66/0.8189/32.19 &32.21/0.8073/31.44 &33.50/0.8343/32.81 &32.61/0.8195/32.00 \\
                        & HAT-S~\cite{hat} &34.56/0.8438/34.53 &35.67/0.8608/35.62 &34.94/0.8505/34.92  &34.74/0.8460/34.71 &35.82/0.8630/35.78 &35.04/0.8529/35.03  \\
                        & SwinIR~\cite{swinir} &34.56/0.8436/34.52 &35.64/0.8605/35.60 &34.93/0.8504/34.90  &34.73/0.8459/34.70 &35.78/0.8627/35.74 &35.02/0.8529/35.01  \\
                        & OAPT(Ours) &\textcolor{red}{34.69}/\textcolor{red}{0.8447}/\textcolor{red}{34.66} &\textcolor{red}{35.78}/\textcolor{red}{0.8616}/\textcolor{red}{35.74} &\textcolor{red}{35.08}/\textcolor{red}{0.8520}/\textcolor{red}{35.05}  &\textcolor{red}{34.85}/\textcolor{red}{0.8469}/\textcolor{red}{34.82} &\textcolor{red}{35.91}/\textcolor{red}{0.8637}/\textcolor{red}{35.88} &\textcolor{red}{35.17}/\textcolor{red}{0.8542}/\textcolor{red}{35.15}  \\
    \hline
    \hline
    \multirow{4}{*}{LIVE1\cite{live1}}
                        & JPEG &28.00/0.8220/26.86 &29.55/0.8649/28.48 &28.61/0.8405/27.84 &28.02/0.8229/26.74 &29.56/0.8655/28.40 &28.58/0.8422/27.57 \\
                        & HAT-S~\cite{hat} &30.20/0.8701/30.01 &31.87/0.9040/31.59 &30.85/0.8848/30.73  &30.44/0.8760/30.16 &32.10/0.9085/31.72 &31.04/0.8904/30.82 \\
                        & SwinIR~\cite{swinir} &30.21/0.8701/30.03 &31.86/0.9039/31.61 &30.87/0.8856/30.77  &30.44/0.8764/30.17 &32.09/0.9086/31.73 &31.07/0.8912/30.86  \\
                        & OAPT(Ours) &\textcolor{red}{30.26}/\textcolor{red}{0.8712}/\textcolor{red}{30.11} &\textcolor{red}{31.92}/\textcolor{red}{0.9050}/\textcolor{red}{31.69} &\textcolor{red}{30.95}/\textcolor{red}{0.8870}/\textcolor{red}{30.84}  &\textcolor{red}{30.51}/\textcolor{red}{0.8773}/\textcolor{red}{30.25} &\textcolor{red}{32.16}/\textcolor{red}{0.9094}/\textcolor{red}{31.80} &\textcolor{red}{31.14}/\textcolor{red}{0.8924}/\textcolor{red}{30.93}  \\
    \hline
    \hline
    \multirow{4}{*}{BSDS500\cite{bsds500}}   
                        & JPEG &28.12/0.8233/26.82 &29.70/0.8685/28.48 &28.75/0.8430/27.83 &28.15/0.8251/26.64 &29.72/0.8695/28.29 &28.74/0.8454/27.49 \\
                        & HAT-S~\cite{hat} &30.20/0.8689/29.96 &31.89/0.9049/31.57 &30.84/0.8840/30.69  &30.42/0.8746/30.04 &32.09/0.9090/31.58 &31.03/0.8897/30.73  \\
                        & SwinIR~\cite{swinir} &30.19/0.8689/29.97 &31.88/0.9048/31.58 &30.87/0.8849/30.72  &30.42/0.8749/30.04 &32.09/0.9093/31.58  &31.06/0.8905/30.76  \\
                        & OAPT(Ours) &\textcolor{red}{30.24}/\textcolor{red}{0.8699}/\textcolor{red}{30.05} &\textcolor{red}{31.92}/\textcolor{red}{0.9056}/\textcolor{red}{31.65} &\textcolor{red}{30.93}/\textcolor{red}{0.8861}/\textcolor{red}{30.79}  &\textcolor{red}{30.47}/\textcolor{red}{0.8757}/\textcolor{red}{30.13} &\textcolor{red}{32.13}/\textcolor{red}{0.9099}/\textcolor{red}{31.66} &\textcolor{red}{31.12}/\textcolor{red}{0.8917}/\textcolor{red}{30.82}  \\
    \hline
  \end{tabular}
  \label{tab:sota_color_table}
}
\end{table}

\subsection{Double JPEG Image Restoration}
As double JPEG image restoration is a relatively new task, we compare our method with DnCNN~\cite{dcnn}, RNAN~\cite{rnan}, SwinIR~\cite{swinir}, HAT~\cite{hat}, ART~\cite{art} and FBCNN~\cite{fbcnn}. As for the experiment of grayscale double JPEG image restoration, except FBCNN, the others are all fine-tuned on the double JPEG compression dataset. For relatively fair comparison, we train the small version of HAT(HAT-S), and use its pretrain weights for super-resolution task to initialize it. Both SwinIR and the reconstructor of our model are initialized on the weights of SwinIR for JPEG artifacts reduction, and we also train ART with its pretrain model for JPEG artifacts reduction. Following the same setting in~\cite{swinir, fbcnn}, the PSNR, SSIM~\cite{ssim} and PSNR-B~\cite{psnrb} results are used as main metrics. 

\begin{figure}[!t]
\centering
\includegraphics[width=\textwidth]{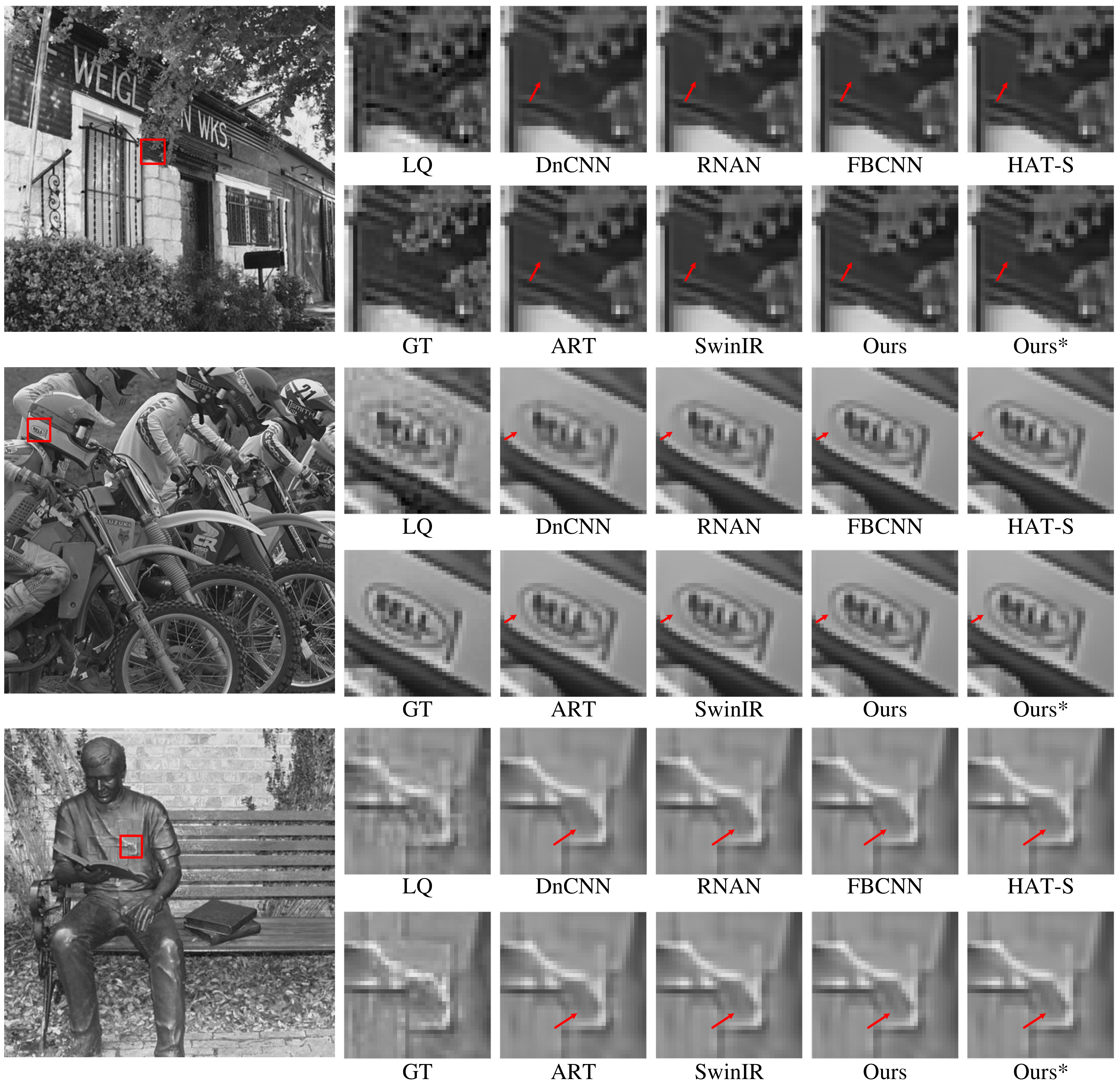}
\caption{Visual comparison for various methods on benchmark datasets. The compression type is $\text{QF}_1=30$, $\text{QF}_2=50$, offsets=$(4, 4)$. }
\label{fig:visual_examples} 

\end{figure}
\begin{figure}[!t]
	\centering
	\includegraphics[width=\textwidth]{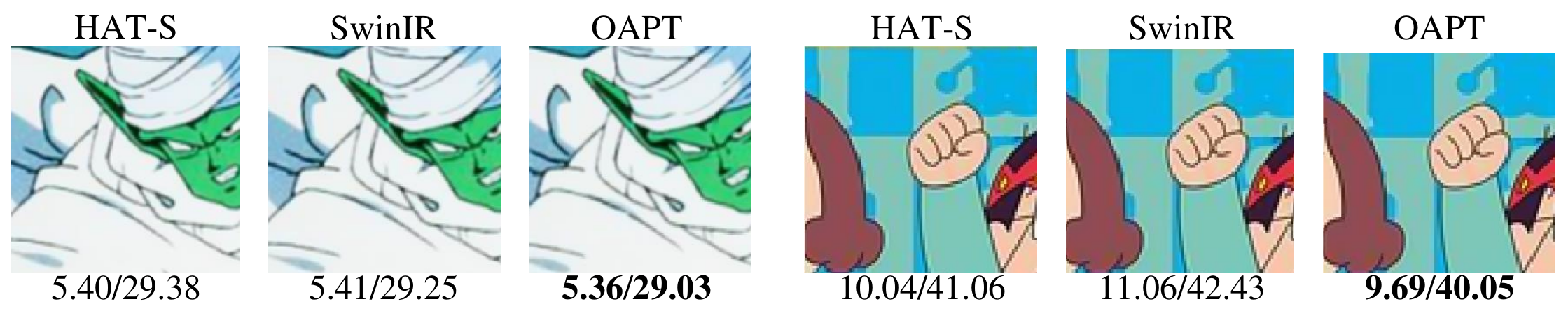}
	\caption{NIQE$\downarrow$ / BRISQUE$\downarrow$ results on real-world images.}
	\label{fig.iqa}
\end{figure}

\begin{table}[ht]
  \centering
  \caption{PSNR/SSIM results on Classic5 and NIQE$\downarrow$ / BRISQUE$\downarrow$ results on RealSRSet of random-compression experiments and PSNR/SSIM results on LIVE1 of singe-compression experiments.}
  \resizebox*{0.8\columnwidth}{!}{
    \begin{tabular}{|c|c|c|c|c|c|}
    \hline
    \multirow{2}{*}{Method}   & \multicolumn{4}{|c|}{Classic5~\cite{class5}(Compression rounds)} & RealSRSet\cite{bsrgan}\\
    \cline{2-6}
    & 2 & 4 & 5 & 7 & blind \\
    \hline
    \hline
    HAT-S~\cite{hat} &32.02/0.8654 &31.25/0.8509 &29.91/0.8174 &28.34/0.7710 &9.89/29.56 \\
    SwinIR~\cite{swinir} &32.04/0.8661 &31.25/0.8508 &29.96/0.8191 &28.29/0.7710 &9.56/29.38 \\
    OAPT(Ours) &\textcolor{red}{32.12}/\textcolor{red}{0.8682} &\textcolor{red}{31.27}/\textcolor{red}{0.8513} &\textcolor{red}{29.96}/\textcolor{red}{0.8195} &\textcolor{red}{28.36}/\textcolor{red}{0.7722} &\textcolor{red}{9.55}/\textcolor{red}{29.27} \\
    \hline
    \hline
    \multirow{2}{*}{Method} & \multicolumn{5}{|c|}{LIVE1~\cite{live1}(QF)} \\
    \cline{2-6}
    & 10 & 20 & 30 & 40 & 50 \\
    \hline
    SwinIR~\cite{swinir} &29.75/0.8245 &32.13/0.8871 &33.53/0.9139 &34.51/0.9288 &35.32/0.9391 \\
    OAPT(Ours) &\textcolor{red}{29.77}/\textcolor{red}{0.8260} &\textcolor{red}{32.16}/\textcolor{red}{0.8886} &\textcolor{red}{33.57}/\textcolor{red}{0.9154} &\textcolor{red}{34.56}/\textcolor{red}{0.9301} &\textcolor{red}{35.38}/\textcolor{red}{0.9403} \\

    \hline
  \end{tabular}
  \label{tab:random_exp}
}
\end{table}

\subsubsection{Quantitative Evaluation.}
We test different methods with eight combinations of QFs and offsets on four benchmark datasets, \ie Classic5~\cite{class5}, LIVE1~\cite{live1}, BSDS500~\cite{bsds500} and ICB~\cite{icb}. The quantitative results for grayscale images are shown in \cref{tab:sota_table}. As observed, the proposed OAPT outperforms most previous methods on PSNR, SSIM and PSNR-B performance. 
Compared with prior state-of-the-art method~\cite{fbcnn}, OAPT achieves great improvement of \textbf{0.16dB} averagely on gray images. 
The result also shows OAPT improves about \textbf{0.18dB} for the hard cases mentioned in FBCNN~\cite{fbcnn} which are non-aligned double JPEG images with $\text{QF}_1\leq\text{QF}_2$.
Besides, compared with SwinIR~\cite{swinir}, OAPT only increases 1.5M parameters of offset predictor but improves averagely about 0.06dB on three datasets. 
The results on color double JPEG images are illustrated in \cref{tab:sota_color_table}. We mainly compare OAPT with SwinIR and HAT-S~\cite{hat}, which have similar parameters with ours.
We observe that our OAPT outperforms SwinIR on all settings and achieves the averagely improvement of 0.08dB over SwinIR. 

\subsubsection{Visual Comparison.}
\cref{fig:visual_examples} illustrates visual quality of two grayscale images in LIVE1~\cite{live1} with type $(\text{QF}_1, \text{QF}_2, i, j)=(30, 50, 4, 4)$. As shown in the examples, OAPT can not only remove the compression artifacts and the non-aligned compression borderlines, but also enrich details of image. Moreover, Ours$^*$, which is OAPT equipped with ground-truth offsets, generates better visual results. 

\subsection{Real-Word JPEG Image Restoration}
To further demonstrate the effectiveness of our OAPT model, we first conduct the random-compression experiment on Classic5~\cite{class5} and RealSRSet~\cite{bsrgan} to verify the univerality of OAPT to multi-compressed image restoration. In \cref{tab:random_exp}, we compressed the images under diverse compression rounds with random QFs and offsets to simply simulate the multi-compression situation, which occurs in the real world. And we also test SwinIR and OAPT on grayscale single-compressed images of LIVE1~\cite{live1} under QF=10, 20, 30, 40, 50. The results shows although the OAPT model is trained with double compression datasets, it has greater robustness on multi-compressed images and single-compressed images.
 
 Also, we conduct experiments on real-world JPEG images. We test models on several pictures collected from the internet, these images were compressed more than once and gone through some other unknown degradations. 
With so many non-reference quality assessment methods~\cite{clipiqa,rlpips,ada-dqa,zoom-vqa,cap-vqa}, we choose the commonly-used NIQE~\cite{niqe} and BRISQUE~\cite{brisque} indexes and obtain the results on the Y channel of these images, and the visual results are shown in \cref{fig.iqa}. It illustrates our model still surpasses other compared methods.

\begin{figure}[t]
	\centering
	\includegraphics[width=0.95\textwidth]{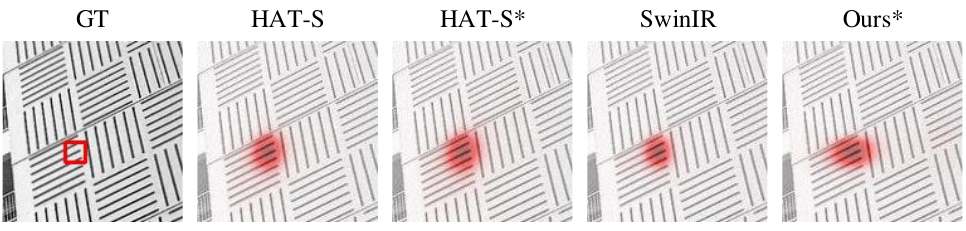}
	\caption{LAM~\cite{lam} results for different methods. * denotes the model equipped with pattern clustering plugin and the ground-truth offsets.}
	\label{fig:lam}
\end{figure}

\begin{figure}[t]
	\centering
	\includegraphics[width=\textwidth]{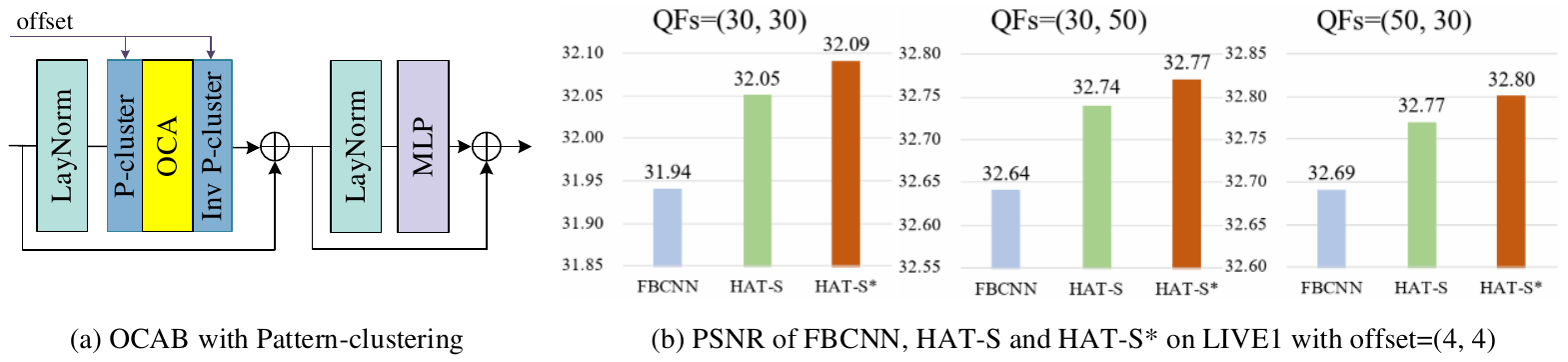}
	\caption{The OCAB module in HAT-S~\cite{hat} with pattern clustering plugin module and PSNR results. }
	\label{fig:hat_with_plugin}
\end{figure}

\subsection{The Effectiveness of the Plugin Module}
\label{sec:exp_of_plugin}
To assess the effectiveness of the pattern clustering plugin module, we strategically insert the plugin module in the Overlapping Cross-Attention Block (OCAB) of HAT-S~\cite{hat} and name the new network as HAT-S$^*$.
\cref{fig:hat_with_plugin}(a) illustrates the integration of OCAB with our plugin module. We refrain from adding an offset predictor to HAT-S and use ground-truth offsets for the pattern clustering module to train the whole network. The results illustrated in \cref{fig:hat_with_plugin}(b), shows HAT-S with this cheap plugin module can easily improve 0.03dB of PSNR value over the base model fine-tuned on double compression dataset averagely without additional parameters and computation complexity.

Furthermore, with the assistance of LAM~\cite{lam}, we can identify pixels that significantly contribute to the selected region of the output image. As illustrated in \cref{fig:lam}, the range of contributing pixels expands in HAT-S when our plugin module is applied. Additionally, Ours$^*$, which is the OAPT model equipped with the ground-truth offset, incorporates more pixels than SwinIR. It affirms that our hybrid attention can not only avoid distraction from degradation of different patterns, but also enlarge the receptive field by non-local grouping.

\subsection{Ablation Study}
\subsubsection{Network structure.}
\begin{figure}[t]
\centering
    \includegraphics[width=0.9\linewidth]{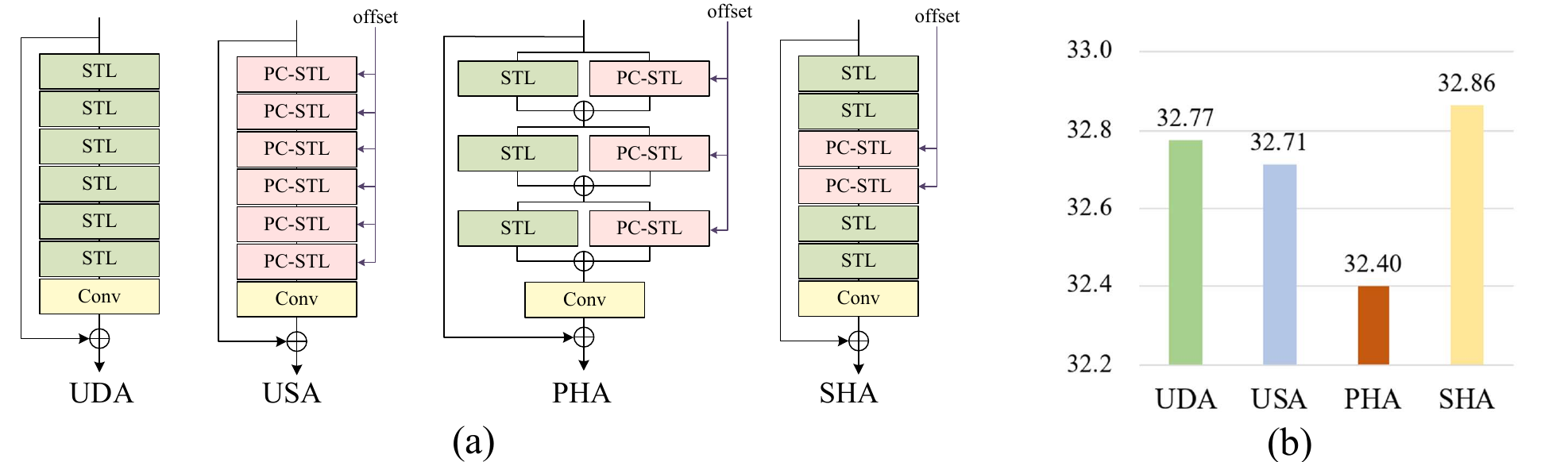}
\caption{Ablation study on various structures of attention.}
\label{fig.HPAB_diff_structures}
\end{figure}

In this section, we evaluate the effectiveness of the HPAB structure. We designed four structures with the same parameters shown in \cref{fig.HPAB_diff_structures}(a): the Serial Hybrid Attention (SHA), the Uniformed Sparse Attention (USA), the Parallel Hybrid Attention (PHA), and the Uniformed Dense Attention (UDA). 
\cref{fig.HPAB_diff_structures}(b) is the PNSR results of networks equipped with different structures under type=$(30, 50, 4, 4)$. It shows all the Serial structures outperform the Parallel one. The parallel structure may introduce confusion in conducting different attentions, thereby compromising the representation of deep features. 
As for the serial structures, the network with SHA structure surpasses that with the USA structure in PSNR results. The USA structure, which destroys the natural image structure and only focuses on non-local features, yields the second least favorable results, highlighting the necessity of dense attention and locality. Furthermore, when the pattern-clustering module is omitted, the SHA structure is simplified to the UDA, which is equivalent to Residual Swin Transformer Block (RSTB) in SwinIR~\cite{swinir}, and it also has less PSNR value than SHA. The result emphasizes the effectiveness of our hybrid attention blocks. In summary, the SHA structure is proved to be relatively more effective than the alternatives.

\begin{figure}[t]
	\centering
	\includegraphics[width=\textwidth]{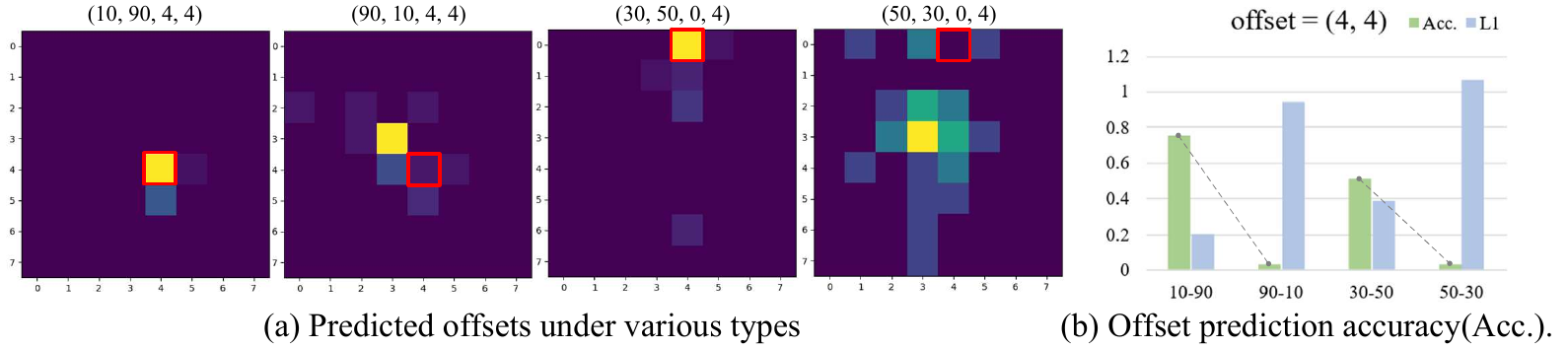}
	\caption{Illustration of prediction results. Fig.9(a) is the distribution map of predicted offsets, the brightness in cells denotes the quantity of predicted offsets, and the coordinates of the cell marked with red lines represents the ground-truth offset. Fig.9(b) is the combined bar graph for results of accuracy and average L1 distance of predicted offsets and ground-truth offsets.}
	\label{fig:predictor}
\end{figure}

\subsubsection{Offset prediction.}
We pretrained the offset predictor first with $44\times44$-sized patches of double compressed images from training datasets. And during the training phase, we freezed the parameters of offset predictor and fine-tuned the image reconstructor with the loss in \cref{rec_loss} for more stable training state. \cref{fig:predictor}(b) denotes the accuracy of offset predictor and L1 distance between predicted offsets and ground-truth offsets under type=$(QF_1, QF_2, 4, 4)$, it shows the accuracy is somehow limited, especially when $QF_1 > QF_2$. While $QF_1 < QF_2$, with the gap between $QF_1$ and $QF_2$ getting bigger, the accuracy gets much higher. We suppose when $QF_1 \geq QF_2$, which means the degree of the second compression is high enough to ruin and cover the blocky boundaries caused by the first compression, the predictor can hardly find any clue to estimate the offsets on the images. Thus, the cases with $QF_1 \geq QF_2$ are \textbf{hard} cases for prediction. Also, in \cref{fig:predictor}(a), we find that the predicted offsets are prone to settling down round the centers of all potential values of offset, like (3, 3), due to the offset loss function in \cref{offset_loss}, when encountering hard cases for prediction. 

To find out whether the ground-truth offsets are significant to the network and how the prediction accuracy affects it, we conducted the experiment as follows. We remove the offset predictor from OAPT and fine-tune it with ground-truth offsets making the OAPT model a non-blind model, termed as OAPT* or Ours*, which shares the same parameters with SwinIR. \cref{tab:predictor} illustrates some results of HAT-S, SwinIR, OAPT and OAPT* under other compression types and the accuracy of the offset predictor in OAPT. We find OAPT gets the best performance averagely with ground-truth offsets (OAPT*) and increases no more parameters and computation complexity, improving about \textbf{0.08dB} over SwinIR on average. Moreover, with the prediction accuracy increasing, the gap of performance between OAPT and OAPT* gets narrower. It indicates with the low accuracy, the performance only benefits from larger receptive field by non-local grouping, while with the high accuracy, it benefits from both larger receptive field and proper pattern clustering by correct offsets for better information extraction.

\begin{table}[t]
  \centering
  \caption{PSNR/SSIM results and the prediction accuracy(Acc.).}
  \resizebox*{0.9\columnwidth}{!}{
    \begin{tabular}{|c|c|c|c|c|c|c|}
    \hline
    Dataset   & Type & HAT-S~\cite{hat} & SwinIR~\cite{swinir} & OAPT(Ours) &OAPT*(Ours*) & Acc. \\
    \hline
    \hline
    \multirow{5}{*}{LIVE1\cite{live1}}
                        & (10, 90, 0, 2) & 29.70/0.8250 & 29.76/0.8259 & \textcolor{red}{29.77}/0.8267 & \textcolor{red}{29.77}/0.8266 & 97\%\\ 
                        &(30, 50, 4, 4) & 32.74/0.9018& 32.77/0.9023 & 32.86/0.9043 & \textcolor{red}{32.87}/0.9046 & 52\%\\
                        & (40, 70, 1, 5) & 33.95/0.9222 & 33.99/0.9227 & 34.07/0.9241 & \textcolor{red}{34.10}/0.9247 & 34\% \\ 
                        &(70, 40, 5, 1) & 29.39/0.8156 & 29.44/0.8169 & 29.48/0.8189 & \textcolor{red}{29.53}/0.8204 & 3\% \\ 
                        &(90, 10, 2, 0) & 29.51/0.8193 & 29.54/0.8200 & 29.57/0.8217 & \textcolor{red}{29.62}/0.8226 & 0\%\\ 
    \hline

  \end{tabular}
  \label{tab:predictor}
}
\end{table}

\section{Conclusion}
In this paper, we propose a method for double compressed image restoration, named as Offset-Aware Partition Transformer(OAPT), which consists of a compression offset predictor and a hybird attention-based image reconstructor. 
Based on the hybrid attention mechanism, OAPT can adjust to the non-aligned double compression by effective pattern clustering operation.
Extensive experiments demonstrate that our method outperforms state-of-the-art  methods.

\textbf{Limitation.}
In our work, the offset predictor is not consistently reliable. Incorrectly predicted offsets merely expand the receptive field, leading to minimal improvement. 
Enhancements can be achieved by increasing prediction accuracy or implicitly utilizing the predicted offsets.


\bibliographystyle{splncs04}
\bibliography{main}

\end{document}